# Letter to the Editor

Against the Resilience of Rejected Manuscripts

Sir,

In a recent Editorial, Cronin (2012) brought to light the topic on the fate of the rejected manuscripts. Indeed, thousands of papers are rejected and resubmitted to other journals with the hope of getting, eventually, published. Although the peer review was originally conceived to filter 'good science' from 'bad science', its shortcomings have been proved many times, such as: the rejection of major discoveries (Anon, 2003) and highly cited papers (Campanario, 2007), the obliviousness of fraud (Armstrong, 1997) and its slow and expensive ineffectiveness (Weller, 2001). It actually works not as a gate keeper of science but as a moderator of science; usually sending manuscripts of lesser quality to journals of lesser impact as it was well pointed out by Cronin (2012). Considering that a submission usually involves one or two editors who have to select at least two reviewers and acknowledging that finding the latter it usually involves contacting around ten researchers (Perrin, 2008); this means that if a single rejected manuscript is then resubmitted to another journal, it will mobilize between three and five people every time it is sent to a journal, increasing in numbers each time it is sent to another journal. But resubmission does not only mean disturbing many colleagues, but also wasting in many cases their efforts, as reviewers' reports from the first journal are not reused by new reviewers and are in fact, thrown away.

Recently, Rohr & Martin (2012) suggested the reuse and recycle of scientific reviews as a way for maximizing reviewers' work and reducing the time lapses between resubmission and editor's answer. The idea behind this would be for editors to request, if previously rejected, other journals' reports on the submitted manuscript. Therefore, editors could check and use these reports in order to decide whether the manuscript is worth reviewing again or not, if the changes made by authors are acceptable or even if the authors' answer to these reports suffices to accept the manuscript or send it to another round of peer review. This proposed system would obviously work against authors as their manuscripts' weaknesses would go unnoticed to the new journals where they'd submit their manuscript. Also, they would be forced to always respond to criticisms. Consequently, some authors would presumably give up their persistence while others would improve their manuscripts. In fact they would carefully consider the journal to which they are submitting their manuscript assuring if their study is within the scope of their target audience. Another advantage would be reducing contradictions between reviewers as the latter would be able to work with the former's reports.

In fact, this approach has already been introduced by some journals (Anon, 2008). However, it also has many shortcomings. Firstly, no journal likes to publish the leftovers of others. In fact, acknowledging that a submitted manuscript has been previously rejected by another journal may well bias the editor's opinion against its acceptance. Although it may occur that a rejected manuscript is accepted in a higher impact journal ending as a highly cited paper (Calcagno et al, 2012), this is a rare exception. Secondly, authors may not wish to send such reports, especially if they are not willing to introduce the suggested changes or to challenge their previous reviewers. Then, the system would remain as it is and no improve would have been made. Journals may not wish to share the reviewers' reports, as some may not like to acknowledge their errors when rejecting manuscripts that end up as major papers (Anon, 2003). And finally, referees themselves may refuse to have their reports used by other journals rather than the one which contact them in the first place.

However, we believe this system would greatly improve the filtering task undertaken by the peer review system. For this purpose, the main editors associations such as the International Committee of Medical Journals, the World Association of Medical Editors or the Council of Science Editors would have to agree and develop clear guidelines. This way a protocol could be developed within online journal management systems, similar to others such as the Open Archives Initiative - Protocol for Metadata Harvesting used in repositories for exchanging metadata or even the ANSI/NISO Z39.50, used in library management systems to interchange bibliographic records worldwide. Such protocol would allow the communication between journals keeping track of rejected manuscripts and their reviews and allowing new journals to reuse them. Although concerns towards possible biases warning against the acceptance of the rejected would still persist, it would undoubtedly release editors and reviewers' workload. Also, it would encourage the authors to rethink before resubmitting, improving the peer review system without involving major changes in the authors' habits.

<em>Accepted for publication in</em> <em>**Journal of the American Society for Information Science and Technology**</em>

**Nicolás Robinson-García**

*EC3: Evaluación de la Ciencia y de la Comunicación Científica,*

*Universidad de Granada,*

*Campus Máximo de Cartuja s/n,*

*18071 Granada*

*Spain*

*E-mail: elrobin@ugr.es*

**Daniel Torres-Salinas**

*EC3: Evaluación de la Ciencia y de la Comunicación Científica,*

*Universidad de Navarra,*

*31008 Pamplona*

*Spain*

*E-mail: torressalinas@gmail.com*

**Juan Miguel Campanario**

*Departamento de Física,*

*Universidad de Alcalá de Henares,*

*28871 Alcalá de Henares, Madrid*

*Spain*

*E-mail: juan.campanario@uah.es*

**Emilio Delgado López-Cózar**

*EC3: Evaluación de la Ciencia y de la Comunicación Científica,*

*Universidad de Granada,*

*Campus Máximo de Cartuja s/n,*

*18071 Granada*

*Spain*

*E-mail: edelgado@ugr.es*